# Coulomb pairing and double-photoionization in aromatic hydrocarbons

D. L. Huber[*]

Physics Department, University of Wisconsin-Madison, Madison, WI 53706, USA

**Abstract**

Recently reported anomalies in the double-photonionization spectra of the aromatic molecules partially deuterated benzene, naphthalene, anthracene, pentacene, azulene, phenanthrene, pyrene and coronene are attributed to Coulomb-pair resonances of $\pi$ electrons. The properties of the resonance in benzene are investigated in detail. The linear behavior in the 2+/1+ ion ratio above the resonance is attributed to a two-electron transition associated with excitation from the ground state to a two-electron continuum. A similar explanation accounts for the linear behavior seen in the pentagonal rings pyrrole, furan, selenophene and thiophene which do not display resonance peaks.

Key Words: aromatic molecules, double-photoionization, coulomb pairing, paired-electron continuum

*Mailing address: Dept. of Physics, University of Wisconsin-Madison, 1150 University Ave., Madison, WI 53711 USA; E-mail address: huber@src.wisc.edu.

## 1. Introduction

Recent studies of double-photoionization in aromatic hydrocarbons have revealed the existence of anomalous peaks in the distribution of doubly charged parent ions in partially deuterated benzene, naphthalene, anthracene, and coronene [1], pentacene [2], azulene and phenanthrene [3], and pyrene [4]. In the case of coronene and pyrene, two peaks were detected, while measurements carried out on the pentagonal rings pyrrole, furan, and selenophene [2] and thiophene [5] do not show a peak. In addition to the peaks, many of the molecules show a linear increase in the ratio of doubly to singly charged ions versus photon energy after subtracting the contribution of the knock-out mechanism. In particular, linear behavior has been seen in partially deuterated benzene, naphthalene, and anthracene with an threshold at the approximate location of the peak, as well as in the pentagonal molecules furan, pyrrole and selenothene [3] and thiophene [5] which do not show a peak. For pyrene and coronene, there is no linear increase beginning at the low energy peak, but there is a linear increase in pyrene at the high energy peak that is not found in coronene [4]. In these papers, it was suggested that the anomalous peaks might be evidence for the existence of bound pairs of electrons analogous to the Cooper pairs associated with superconductivity [6,7].

The purpose of this note is to outline a qualitative interpretation of the anomalous peaks and the linear increases. The interpretation of the peaks is based on the existence of two-body bound states of interacting particles in a periodic potential. It was shown some time ago by Slater *et al* [8] and Hubbard [9] that bound states exist for screened Coulomb interactions between electrons in one-dimensional metals with periodic potentials and short range repulsive interactions. In [10], an analysis is made of the two-electron bound states in a nearest-neighbor tight-binding chain with on-site and nearest-neighbor interactions, while a general investigation of the formation of bound states for finite range repulsive forces is presented in [11]. It should be noted that the authors of [11], who refer to the bound states as 'Coulomb pairs', suggested that benzene might be a system where coulomb pairing could be observed

## 2. Coulomb bound states

In the analysis of the Coulomb bound states we make use of the tight-binding results obtained in [10, 12]. In the tight-binding approximation, which is appropriate when the basis states are Hückel orbitals, the Hamiltonian involves the transfer integral and two additional parameters associated with the Coulomb interaction: the Hubbard contact parameter [9],$U$, which we take to be positive corresponding to on-site Coulomb repulsion, and the nearest-neighbor Coulomb parameter, $J$, which we will also take to be positive. We further assume $U >> J$. Because the bound state is comprised of two electrons, there are two types of spin states that are consistent with the overall antisymmetry of the two-particle wave function: the singlet spin state, $S = 0$, with a symmetric spatial wave function and the triplet spin states, $S = 1$, with an antisymmetric spatial wave function.

When $U \gg J$, the energy of the $S = 0$ (spin singlet) two-electron bound state relative to twice the HOMO-LUMO mid-gap energy is expressed as [10]

$$E_B^{S=0} = (U^2 + 16\beta^2)^{1/2}, \qquad (1)$$

where $\beta$ denotes the transfer integral. In the case of the $S = 1$ (spin triplet) bound state, the $U$ parameter does not enter since the wave function vanishes when the particles occupy the same site. The bound state energy for this system is expressed as

$$E_B^{S=1} = J + (4\beta^2 / J). \qquad (2)$$

As pointed out in [10], the paired and the unoccupied independent (Hückel) states overlap (and thus hybridize) when $4\beta > J$. Also, since $U \gg J$, we have the inequality

$$E_B^{S=0} \gg E_B^{S=1}. \qquad (3)$$

It must be emphasized that the rigorous analysis of the Coulomb bound state applies only to two-electron systems. For this reason, it is more appropriate to speak of a Coulomb-pair resonance rather than a true bound state in the aromatic hydrocarbons, since the pair state rapidly decays into two free electrons.

## 3. Double-photoionization in benzene

As shown in [1-4], double-photoionization measurements carried out on partially deuterated benzene reveal a peak in the ratio of doubly charged parent ions to singly charged ions (after subtracting the contribution from the knock-out mechanism) vs photon energy. We argue the peak is a resonance related to the formation of a localized Coulomb pair comprised of two electrons from the occupied $\pi$-orbitals. In the case of benzene, the simplest theory has the remaining 4 electrons in their initial $\pi$-orbital states. The resonant energy, 42 eV, which is identified with the difference between the continuum threshold and the double ionization threshold [13] (see below), is consistent with a $S = 0$ resonance. Given the value of $\beta$, we can use Eq.(1) to obtain the value of $U$. In the case of benzene, the HOMO-LUMO gap is equal to $2\beta$. Expressing the gap as the difference between the one-electron ionization energy and the electron affinity [14] leads to $\beta = 5.2$ eV. Setting $E_B^{S=0} = 42$ eV, we obtain $U = 36$ eV.

In [8], it is pointed out that the parameter $U$ can be calculated from the Wannier functions associated with the $\pi$-orbitals. In the narrow-band limit $U \gg 4\beta$, the calculation for benzene reduces to the evaluation of the Coulomb energy of a pair of $2p$ electrons on the same carbon atom. This takes the form

$$U = e^2 \int d\mathbf{r} \int d\mathbf{r}' \frac{\rho_{2p}(\mathbf{r})\rho_{2p}(\mathbf{r}')}{|\mathbf{r} - \mathbf{r}'|} \qquad (4)$$

where $\rho_{2p}$ denotes the probability density of the 2p electron. We have evaluated (7) using Slater-type integrals for the radial functions [15]. This approach leads to the result $U \approx 17$ eV. Although smaller than the value inferred from the experimental data, 36 eV, it is in 'order of magnitude' agreement despite the fact that benzene appears to be well outside the narrow-band limit.

The tight-binding approach of Ref. [10] also provides an estimate of the exponential decay rate of the localized state, $\gamma$, which takes the form

$$\gamma = \operatorname{arcsinh}(U/4\beta) = 1.32, \qquad (5)$$

using parameters appropriate to benzene. The exponential decay rate is defined terms of the decrease in amplitude of the localized state wave function at a site which has $m$ intervening bonds between it and the site of maximum amplitude. Thus if the maximum amplitude is $A$, the amplitude at the site three bonds away ($m = 3$) would be $\exp[-3 \times 1.32]A = 0.02A$. The oscillator strength associated with the formation of a Coulomb pair may be enhanced if the ground state of the molecule is characterized by the presence of fluctuating electron pairs [6,7].

The linear behavior above the peak that is seen in benzene and other molecules can be explained if we assume that the benzene peak is located at the threshold of a two-electron continuum. When the photon energy exceeds the energy necessary to form a localized Coulomb pair, two electrons are excited into a continuum state that ultimately decays into two free electrons and a 2+ ion. The sum of the kinetic energy of the electrons is equal to $E_{ph} - E_{thr}$ where $E_{ph}$ is the photon energy and $E_{thr}$ is the energy of the threshold of the continuum. Under steady-state conditions, the 2+ ion concentration is equal to ratio of the formation rate to the recombination rate. The rate of formation of the 2+ ions is proportional to the density of states of the two-electron system. We assume that under steady-state conditions the electrons are uncorrelated so that the two-electron density of states is proportional to the product of the densities of states of the two particles. Assuming the latter are vary as $E^{1/2}$ we can write the density of states of the two-electron system, $DOS(2e)$, as

$$DOS(2e) = A[(E_{ph} - E_{thr})/2], \qquad (6)$$

when the two electrons have the same energy, which is to be expected when the linear behavior is preceded by a pairing resonance. If the two electrons have different energies but the same total energy, $E_{ph} - E_{thr}$, as could happen if there is no pairing resonance, the $DOS(2e)$ can be expressed as an integral:

$$DOS(2e) = A(E_{ph} - E_{thr}) \int_0^1 dx f(x)(1-x)^{1/2} x^{1/2}, \qquad (7)$$

where $A$ is a constant in both (5) and (6), and $f(x)$ is a normalized weighting function, symmetric about $x = ½$, that is equal to $\delta(x-1/2)$ if the two electrons have the same energy. In the opposite limit, where the individual energies are not correlated, $f(x) = 1$, and the integral is equal to $\pi/8$. Note that in both equations, $DOS(2e)$ is proportional to $E_{ph} - E_{thr}$. Assuming the density of single ions, $M(+)$, varies slowly in the neighborhood of the neighborhood of the pair continuum, the measured values of $[M(2+)/M(1+)$ – 'knock-out'] will vary as $E_{ph} - E_{thr}$, which is what was found in the experiments.

## 4. Discussion

The anomalous high-energy peak in the concentration of doubly ionized aromatic molecules has been reported for deuterated benzene, coronene, naphthalene, anthracene, pentacene and pyrene, along with the isomers azulene and phenanthrene. It is likely that the peak is associated with a $S = 0$ resonance. As mentioned, coronene and pyrene show a second peak at about 10 eV relative to the double ionization threshold. It is possible that this peak is associated with a $S = 1$ resonance where the energy is given by Eq. (2). The absence of the low energy peak in the other compounds where there is a high-energy peak could be due to the hybridization of the $S = 1$ localized state with the molecular orbital states. The absence of a peak structure altogether in pyrrole, furan, selenophene and thiophene is attributed to the presence of a non-carbon atom in the pentagonal ring interrupting the periodicity. This attribution is supported by recent measurements on the hexagonal C-N molecues 1-3-5 triazine (C-N-C-N-C-N), pyrimidine (C-N-C-N-C-C) and pyridizine (C-N-C-C-C-C). The resonance in triazine, which has three-fold point symmetry, is nearly the same magnitude as in the six-fold symmetric benzene, while the corresponding resonance in the asymmetric pyrimidine is significantly smaller, and the resonance in the single-impurity pyridizine is barely measurable [13]. The molecule pyrazine (C-C-N-C-C-N) has two-fold symmetry and may also have a large resonant peak.

A common feature of the compounds where the peak has been detected is the presence of benzene-like hexagonal units which is consistent with the localized state being largely confined to a primary site and its near-neighbors as in benzene. The presence of the anomaly in the highly asymmetric azulene, where there are only singlet orbitals, suggests that electrons in both singlet and doublet orbitals can play a role in the higher symmetry molecules. It is likely that contributions from different orbitals are the primary cause of the width of the anomalous peak, which is approximately 10 eV in benzene (HWHM, low energy side) [4], and the two-peak structure associated with the 50 eV resonance in coronene [4].

The hypothesis for the linear behavior of the 2+/1+ ion ratio involves continuum states for electron pairs. It explains the absence of linear behavior associated with the low energy peaks in coronene and pyrene because the peaks lie well below the continuum. In the case of coronene, it could also explain the absence of linear behavior above the high energy peak if there are one or more additional resonance states with higher energy and thus a higher $E_{thr}$. The continuum threshold hypothesis also accounts for the presence of linear behavior in furan, pyrrole, selenophene and thiophene where there are no resonant peaks.

**Acknowledgment**

The author would like to thank Ralf Wehlitz for preprints, references and many helpful comments.